# Pendulum Waves on the Surface of Block Rock Mass under Dynamic Impact

**N. I. Aleksandrova**

*Chinakal Institute of Mining, Siberian Branch, Russian Academy of Sciences,
Krasnyi pr. 54, Novosibirsk, 630091 Russia e-mail: nialex@misd.ru*

**Abstract**—Under numerical investigation is propagation of surface pendulum waves in 3D block medium. The medium is modeled by 3D lattice of masses connected with elastic springs and viscous dampers. The surface vertical pulsed concentrated loading is considered. The displacements and velocities of the surface masses are calculated. The numerical results obtained for the block medium are compared with the similar data on elastic medium and in situ experiments carried out by other researchers.

*Keywords:* Lamb's problem, block medium, half-space, wave motion, Rayleigh wave, numerical modeling.

**DOI:** 10.1134/S1062739117011847

## INTRODUCTION

The fundamental concept put forward by Sadovsky [1], according to which a rock mass is a system of different scale nested blocks parted by interlayers composed of softer jointy rocks, greatly advanced understanding of the process of wave propagation in a block-structured medium. The block structure of a medium is a cause of different dynamic events that are absent in a uniform medium and, thus, undescribable with the matchable models [2]. Among such dynamic events, of interest is travel of pendulum waves that feature low velocity, long length and weak attenuation [2, 3].

A simplest way of studying dynamics of a block-structure medium its pendulum-type approximation when all blocks are assumed incompressible and deformation and displacement of blocks take place due to compressibility of interlayers. The analytical model in this case is a lattice of masses interconnected by springs and dampers. Despite the apparently limited usability of lattices as simulations of real block-structured media, they benefit from the ability to involve analytical and numerical methods and the capacity of qualitative description of dynamic events in such media.

In this study, a block-structure medium is simulated by 3D lattice of masses connected axially and diagonally by springs and dampers. In the framework of the model, Lamb's problem is solved using the finite difference method and the result is compared with the full-scale testing data [4] and with the analytical solutions for a uniform elastic medium [5].

## 1. PROBLEM FORMULATION

We study a nonstationary three-dimensional Lamb's problem on vertical point load applied to the surface of a block-structured half-space. The block-structure medium is simulated with a uniform 3D lattice of point masses interconnected by springs and dampers in the directions of the axes $x$, $y$, $z$ and in the diagonal directions of the planes $x$ = const, $y$ = const, $z$ = const a shown in Fig. 1a. Here, $u$, $v$ — horizontal displacements in the lines of $x$, $y$; $w$ — vertical displacements in the line of $z$; $n$, $m$, $k$ — numbers of blocks in the lines of $x$, $y$, $z$. Let $l$ be the length of the springs along the axes $x$, $y$, $z$. The theoretical description of deformation of the interlayers uses the Kelvin–Voigt rheological model [6]. To the surface of the block-structure half-space, $k = 0$, at the point $O$ with the "coordinates" $n = 0$, $m = 0$, a vertical point load $Q$ is applied (Fig. 1b).

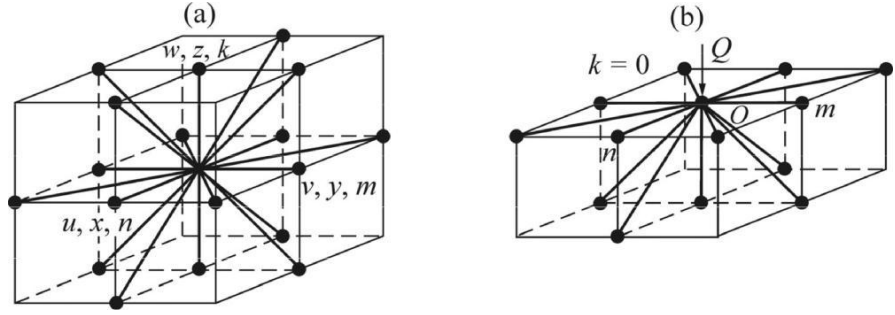

**Fig. 1.** Scheme of connection of masses, springs and dampers (a) inside the half-space and (b) on the surface of the half-space.

The equations of motion of a mass with the coordinates $n, m, k$ inside the half-space are given by

$$M\ddot{u}_{n,m,k} = k_1 \Lambda_{nn} u_{n,m,k} + k_2[(\Phi_{nk} + \Phi_{nm})u_{n,m,k} + \Psi_{nm} v_{n,m,k} + \Psi_{nk} w_{n,m,k}]/2$$
$$+ \lambda_1 \Lambda_{nn} \dot{u}_{n,m,k} + \lambda_2[(\Phi_{nk} + \Phi_{nm})\dot{u}_{n,m,k} + \Psi_{nm} \dot{v}_{n,m,k} + \Psi_{nk} \dot{w}_{n,m,k}]/2;$$
$$M\ddot{v}_{n,m,k} = k_1 \Lambda_{mm} v_{n,m,k} + k_2[\Psi_{nm} u_{n,m,k} + (\Phi_{mk} + \Phi_{nm})v_{n,m,k} + \Psi_{mk} w_{n,m,k}]/2$$
$$+ \lambda_1 \Lambda_{mm} \dot{v}_{n,m,k} + \lambda_2[\Psi_{nm} \dot{u}_{n,m,k} + (\Phi_{mk} + \Phi_{nm})\dot{v}_{n,m,k} + \Psi_{mk} \dot{w}_{n,m,k}]/2; \quad (1)$$
$$M\ddot{w}_{n,m,k} = k_1 \Lambda_{kk} w_{n,m,k} + k_2[\Psi_{nk} u_{n,m,k} + \Psi_{mk} v_{n,m,k} + (\Phi_{mk} + \Phi_{nk})w_{n,m,k}]/2$$
$$+ \lambda_1 \Lambda_{kk} \dot{w}_{n,m,k} + \lambda_2[\Psi_{nk} \dot{u}_{n,m,k} + \Psi_{mk} \dot{v}_{n,m,k} + (\Phi_{mk} + \Phi_{nk})\dot{w}_{n,m,k}]/2,$$

where

$$\Lambda_{nn} f_{n,m,k} = f_{n+1,m,k} - 2f_{n,m,k} + f_{n-1,m,k};$$
$$\Phi_{nm} f_{n,m,k} = f_{n+1,m+1,k} + f_{n-1,m-1,k} - 4f_{n,m,k} + f_{n+1,m-1,k} + f_{n-1,m+1,k};$$
$$\Psi_{nm} f_{n,m,k} = f_{n+1,m+1,k} + f_{n-1,m-1,k} - f_{n+1,m-1,k} - f_{n-1,m+1,k}.$$

The equations of motion of a mass with the coordinates $n, m, 0$ on the surface of the half-space are given by:

$$M\ddot{u}_{n,m,0} = k_1 \Lambda_{nn} u_{n,m,0} + k_2[(\Phi^-_{nk} + \Phi^-_{nm})u_{n,m,0} + \Psi^-_{nm} v_{n,m,0} + \Psi^-_{nk} w_{n,m,0}]/2$$
$$+ \lambda_1 \Lambda_{nn} \dot{u}_{n,m,0} + \lambda_2[(\Phi^-_{nk} + \Phi^-_{nm})\dot{u}_{n,m,0} + \Psi^-_{nm} \dot{v}_{n,m,0} + \Psi^-_{nk} \dot{w}_{n,m,0}]/2;$$
$$M\ddot{v}_{n,m,0} = k_1 \Lambda_{mm} v_{n,m,0} + k_2[\Psi^-_{nm} u_{n,m,0} + (\Phi^-_{mk} + \Phi^-_{nm})v_{n,m,0} + \Psi^-_{mk} w_{n,m,0}]/2 \quad (2)$$
$$+ \lambda_1 \Lambda_{mm} \dot{v}_{n,m,0} + \lambda_2[\Psi^-_{nm} \dot{u}_{n,m,0} + (\Phi^-_{mk} + \Phi^-_{nm})\dot{v}_{n,m,0} + \Psi^-_{mk} \dot{w}_{n,m,0}]/2;$$
$$M\ddot{w}_{n,m,0} = k_1 \Lambda^-_k w_{n,m,0} + k_2[\Psi^-_{nk} u_{n,m,0} + \Psi^-_{mk} v_{n,m,0} + (\Phi^-_{mk} + \Phi^-_{nk})w_{n,m,0}]/2$$
$$+ \lambda_1 \Lambda^-_k \dot{w}_{n,m,0} + \lambda_2[\Psi^-_{nk} \dot{u}_{n,m,0} + \Psi^-_{mk} \dot{v}_{n,m,0} + (\Phi^-_{mk} + \Phi^-_{nk})\dot{w}_{n,m,0}]/2 + Q(t)\delta_{n0}\delta_{m0},$$

where

$$\Lambda^-_k f_{n,m,0} = f_{n+1,m,-1} - f_{n,m,0};$$
$$\Phi^-_{nk} f_{n,m,0} = f_{n-1,m,-1} - 2f_{n,m,0} + f_{n+1,m,-1}; \quad \Phi^-_{mk} f_{n,m,0} = f_{n,m-1,-1} - 2f_{n,m,0} + f_{n,m+1,-1};$$
$$\Psi^-_{nk} f_{n,m,0} = f_{n-1,m,-1} - f_{n+1,m,-1}; \quad \Psi^-_{mk} f_{n,m,0} = f_{n,m-1,-1} - f_{n,m+1,-1}.$$

The notions in (1) and (2) are: $k_1$, $\lambda_1$ — stiffnesses of the springs and viscosities of the dampers in the directions of the axes $x$, $y$, $z$; $k_2$, $\lambda_2$ — spring stiffnesses and damper viscosities in the diagonal directions; $\delta_{n0}$ — Kronecker delta; $Q(t)$ — load. The initial conditions for Eqs. (1) and (2) are zero. The process of derivation of (1) and (2) is described in [7]. Owing to the symmetry, we only discuss the process of wave propagation in the domain $n \geq 0$, $m \geq 0$.

From now on, it is assumed that $k_1 = k_2$ and $\lambda_1 = \lambda_2$. The analysis of dispersion properties in [7] shows that the phase velocities of the longitudinal $c_p$ and shear $c_s$ infinitely long waves are defined by the formulas:

$$q_x \to 0, \quad q_y \to 0, \quad q_z \to 0, \quad c_p = l\sqrt{\frac{3k_1}{M}}, \quad c_s = l\sqrt{\frac{k_1}{M}}. \tag{3}$$

According to the calculations [7], the moduli of the phase and group velocities of long waves in a block-structured medium match. For this reason, it is inferred that the infinitely long waves propagate in a block-structured medium without dispersion and generate low-frequency pendulum waves.

Using Rayleigh's equation for an elastic medium [8] and the formula (3), we calculate the velocity of infinitely long Rayleigh's waves in a block-structured medium:

$$c_R = l\sqrt{\frac{2k_1}{M}\left(1 - \frac{1}{\sqrt{3}}\right)}. \tag{4}$$

## 2. CALCULATION DATA. PULSED LOAD

The equations (1) and (2) with the zero initial conditions were solved by the explicit-scheme finite-difference method. The approximations of derivatives with respect to time, which were used in the calculations, are given below:

$$\ddot{u}_{n,m,k} \approx \frac{u^{s+1}_{n,m,k} - 2u^s_{n,m,k} + u^{s-1}_{n,m,k}}{\tau^2}, \quad \dot{u}_{n,m,k} \approx \frac{u^s_{n,m,k} - u^{s-1}_{n,m,k}}{\tau}, \quad k \leq 0, \quad s = 0, 1, 2,...$$

Here, $\tau$ — time step of the difference scheme; $u^s_{n,m,k}$ — value of the displacement $u_{n,m,k}(t)$ at the time moment $t = s\tau$; $s$ — number of the time layer in the finite difference scheme. The similar approximations are used for the displacements $v_{n,m,k}(t)$, $w_{n,m,k}(t)$. The condition of stability of the difference equations in Lamb's problem when $\lambda_1 = 0$ is given by: $\tau \leq l\sqrt{M/3k_1}$. Let us discuss Lamb's problem for a block-structured medium exposed to vertical pulsed loading modeled by a half-sinusoid with a duration $t_0$:

$$Q(t) = Q_0 \sin(\omega t) H(t) H(t_0 - t), \quad \omega = \frac{\pi}{t_0}, \tag{5}$$

where $H(t)$ — Heaviside step function; $Q_0$ — load amplitude.

Figure 2 shows the radial $u_r = (nu_{n,m,0} + mv_{n,m,0})/\sqrt{n^2 + m^2}$ and vertical $w = w_{n,m,0}$ displacements in the cylinder coordinates $r$, $\theta$, $z$ and their velocities $\dot{u}_r$, $\dot{w}$ as functions of time, numerically — for a block-structured medium at the point $n = 30$, $m = 0$ and analytically elastic medium with Poisson's ratio $\sigma = 0.25$ at the point $r = 30$ [5]. For the both media assumed that $\omega = 0.15$. All calculations using Eqs. (1), (2) involved: $Q_0 = l = M = 1$, $\lambda_1 = 0$, $\tau = 0.5$. The vertical lines in Fig. 2 conform with the arrivals of the quasi-fronts of P-, S- and Rayleigh's waves at the pre-assigned point $(n, 0, 0)$:

$t_p = n/c_p$, $t_s = n/c_s$, $t_R = n/c_R$, where $c_p, c_s, c_R$ are found from (3), (4). The vertical line at $t_R + t_0$ corresponds to the tail front of Rayleigh's wave, which lags behind its head front by the time equal to the active pulse duration.

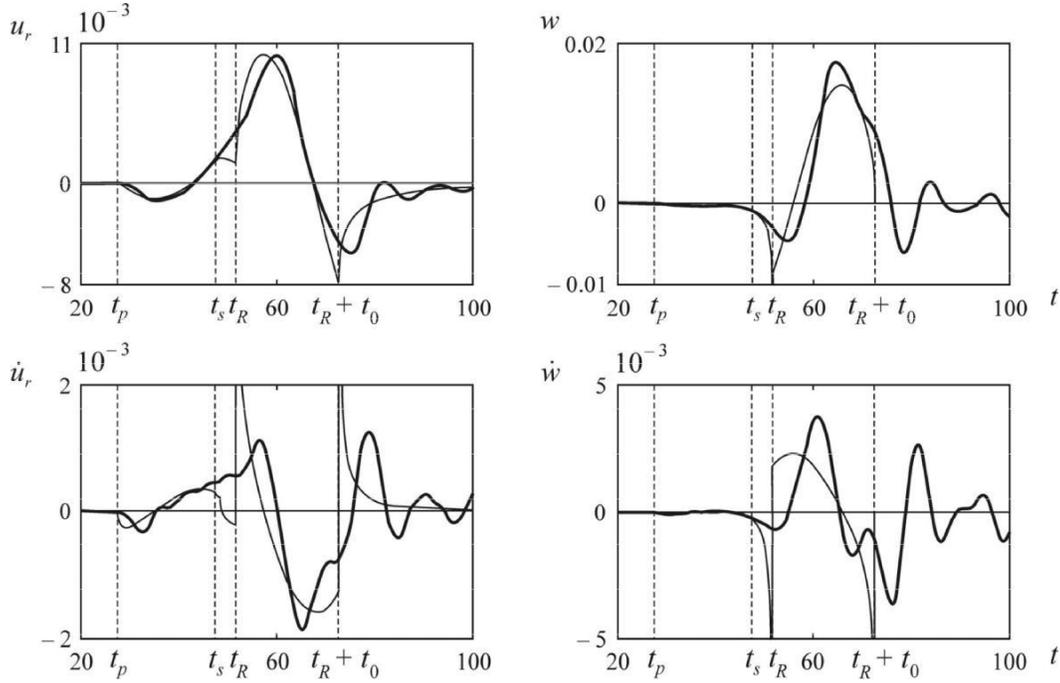

**Fig. 2.** The radial displacements $u_r$, vertical displacements $w$ and their velocities $\dot{u}_r$, $\dot{w}$, calculated for a block-structure medium (thick lines) and for an elastic medium (thin lines).

The support e-materials to [5] present a calculation program for the vertical and radial displacements on the surface of an elastic half-space under a stepped load. We, first, calculated the vertical and radial displacements, $u_r$ and $w$, on the surface of the elastic half-space using the mentioned program. Then, by the numerical differentiation of the displacements with respect to time, we computed the vertical and radial displacement velocities $\dot{u}_r$ and $\dot{w}$ at the same point. The perturbation of the elastic medium under the pulsed load was calculated using Duhamel's integral:

$$f(t) = \int_0^t g(\tau) \frac{\partial Q(t-\tau)}{\partial t} d\tau,$$

where $g(\tau)$ — response of the medium to the stepped load; $f(t)$ — response of the medium to the pulsed load (5).

It is seen that the solutions for the block-structure and elastic media differ. Nonetheless, there is qualitative conformity, namely, the block-structure medium solution fluctuates nearby the elastic medium solution.

### 3. COMPARISON OF THE EXPERIMENTAL AND THEORETICAL DATA

Aiming to find the capacity of the block-structure model (1), (2) to describe propagation of pendulum waves in real rock masses, we compare the full-scale experimental and theoretical results.

The case study [4] reports experimental data obtained at an open pit limestone quarry of Iskitimizvest company in the Novosibirsk Region. In the tests, seismic waves generated by impacts were recorded on the surface of rock blocks, as well the velocities, amplitudes and spectra of the waves were determined at various distances from the impact point. The researchers [4] mentioned that in a nonfractured limestone specimen taken from the quarry, P-waves traveled at the velocity of 5.5 km/s. An example of the seismograms of the displacement velocity from [4] is given in Fig. 3. The impact starts at $t = 0.045$ s. Each seismogram is normalized to have the same amplitude. The straight line *1* in Fig. 3 indicates the time of the first perturbations at that point. The slope of the line conforms with the travel velocity of 3.7 km/s. the straight line *2* points at the time of the first wave with the maximum amplitude. The slope of this line conforms with the wave velocity of 2.3 km/s. Thus, according to these data, P-wave velocity in the block-structured medium (3.7 km/s) is much lower than P-wave velocity in the materials free from jointing (5.5 km/s).

Figure 4 gives the calculated data for the radial displacement velocity $\dot{u}_r$, on the surface of the half-space from 3D block-structure medium modeling (1), (2) under the influence of the vertical pulsed load (5) applied at the point (0, 0, 0). The problem parameters have the values: $\omega = 0.15$, $Q_0 = l = M = 1$, $k_1 = 0.4$, $\lambda_1 = 0.01$, $\tau = 0.5$. the radial displacement velocities $\dot{u}_r$, are calculated at the points $n = 10, 16, 22, 28, 34, 40$, $m = k = 0$. Each curve in Fig. 4 is normalized in the same manner as the experimental seismograms in Fig. 3. The straight lines *1* and *3* in Fig. 4 intersect the curves at the points that conform with the time moments $t = n/c_p$ and $t = n/c_R$ of arrival of P- and Rayleigh's waves at the point ($n$, 0, 0). The straight line *2* is drawn in parallel to *3* through the points $t = n/c_R + t_0$. It is evident in Fig. 4 that the lag between the Rayleigh wave and the next wave is equal to the duration of the active pulse (5).

This study is not aimed to find correspondence between the physical characteristics of limestone and the dimensionless parameters of the theoretical model of a block-structure medium, which were used in the calculations in Fig. 4. Our goal is to demonstrate the qualitative conformance of the experimental seismograms (Fig. 3) and the calculated data (Fig. 4). The signs of the qualitative conformance are: the seismograms and the theoretical curves behave similarly; the straight lines *1* and *2* situate relative to the seismograms and the theoretical curves in a resembling way; the radial velocity fluctuates after the time moment $t = n/c_R + t_0$, which is absent in case of the uniform elastic medium (Fig. 2). Based on that, the author draws a conclusion that the block-structure medium model (1), (2) describes the experimental results much better than the classical model of the uniform isotropic elastic medium.

## CONCLUSIONS

The author has calculated propagation of seismic waves using the three-dimensional mathematical model of block-structured rock mass based on the idea that the dynamic behavior of a block medium may be approximately described as the displacement of rigid blocks owing to the yield of the interlayers between the blocks, while the deformation of the interlayers may be approximated by the Kelvin–Voigt model. The numerical data and the actual testing results conform qualitatively.

The research findings imply that: the calculations of seismic waves should account for the block structure of real rock masses and for the rheological properties of the block interlayers; the three-dimensional block-structure medium model (1), (2) offers the qualitatively correct description of the seismic wave travel in rocks.

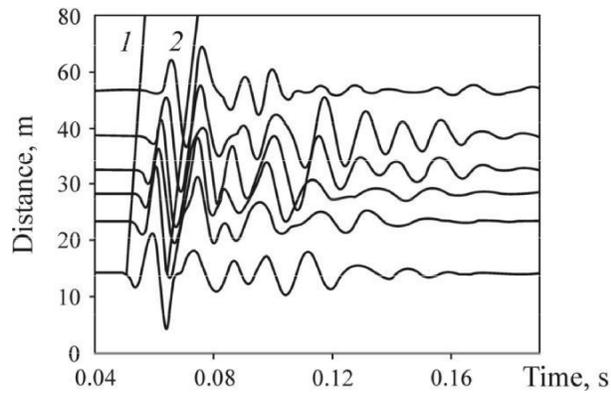 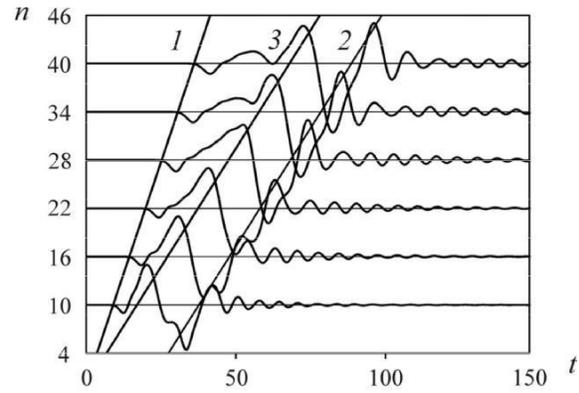

**Fig. 3.** Seismograms of the radial displacement velocities on the surface of blocks subjected to impact loading, recorded at different distances from the impact point: taken from [4].

**Fig. 4.** The radial displacement velocity $\dot{u}_r$, on the surface of the half-space, calculated from the three-dimensional model (1), (2) of a block-structure medium under pulse loading.

ACKNOWLEDGMENTS

This study was supported in the framework of project ONZ RAN-3.1.